\documentclass[fleqn,10pt]{wlscirep}
\usepackage[utf8]{inputenc}
\usepackage[T1]{fontenc}
\usepackage{lineno}
\usepackage[version=4]{mhchem}
\usepackage{textcomp, gensymb}
\usepackage{booktabs} 
\usepackage{array}
\usepackage{xr}
\title{Benchmarking Autonomy in Scientific Experiments: A Hierarchical Taxonomy for Autonomous Large-Scale Facilities}

\author[1,2,3,*]{James Le Houx}

\affil[1]{University of Greenwich, Old Royal Naval College, Park Row, London, SE10 9LS, United Kingdom}
\affil[2]{ISIS Neutron \& Muon Source, Rutherford Appleton Laboratory, Didcot, OX11 0QX, United Kingdom}
\affil[3]{The Faraday Institution, Harwell Science and Innovation Campus, Didcot, OX11 0RA, United Kingdom}

\affil[*]{corresponding author: James Le Houx (james.le-houx@stfc.ac.uk)}

\begin{abstract}
The transition from automated data collection to fully autonomous discovery requires a shared vocabulary to benchmark progress. While the automotive industry relies on the SAE J3016 standard, current taxonomies for autonomous science presuppose an owner-operator model that is incompatible with the operational rigidities of Large-Scale User Facilities. Here, we propose the Benchmarking Autonomy in Scientific Experiments (BASE) Scale, a 6-level taxonomy (Levels 0--5) specifically adapted for these unique constraints. Unlike owner-operator models, User Facilities require zero-shot deployment where agents must operate immediately without extensive training periods. We define the specific technical requirements for each tier, identifying the Inference Barrier (Level 3) as the critical latency threshold where decisions shift from scalar feedback to semantic digital twins. Fundamentally, this level extends the decision manifold from spatial exploration to temporal gating, enabling the agent to synchronise acquisition with the onset of transient physical events. By establishing these operational definitions, the BASE Scale provides facility directors, funding bodies, and beamline scientists with a standardised metric to assess risk, define liability, and quantify the intelligence of experimental workflows.\\

\textbf{Keywords:} Autonomous Experimentation, Large-Scale User Facilities, AI for Science, Taxonomy, Operational Design Domains (ODD), Sim-to-Real Transfer

\end{abstract}
\begin{document}
\flushbottom
\maketitle

\thispagestyle{empty}

\section*{The Data-Insight Gap in Large-Scale Facilities}

Large-scale facilities have historically prioritised data velocity over information density. Although the community has successfully automated the mechanical collection of data, the industrialisation of insight extraction remains unresolved. As Drnec and Lyonnard highlight, the field faces a crisis in Reliability, Representativeness, and Reproducibility~\cite{drnec2025battery}. This bottleneck is not unique to large-scale facility science; it parallels challenges in fields such as adaptive magnetic resonance imaging (MRI)~\cite{song2009optimal} and neuromorphic engineering~\cite{chen2020event}, where data acquisition is increasingly gated by information content rather than fixed clock cycles. Consequently, current high-throughput workflows generate massive datasets that statistically miss the critical transient events~\cite{lu2025autonomousbatteryresearchprinciples}.

Current brute-force automation strategies are inherently inefficient, as a standard grid scan wastes over 99\% of available beam flux measuring the stable bulk of a material. Consequently, it misses the rare localised plasticity that governs performance. Recent work on high-nickel cathodes demonstrates that failure is governed by deterministic geometric features~\cite{LeHoux2026}. A scripted automation system cannot distinguish these information-rich features from the bulk. True efficiency therefore requires an experimental logic that actively hunts for these features to maximise the Entropy-Scaled Measurement Efficiency ($E_{\text{SME}}$)~\cite{lu2025autonomousbatteryresearchprinciples}.

Despite this clear scientific imperative, the transition from automated execution to autonomous discovery is currently hindered by semantic ambiguity. Terms such as AI-driven and autonomous are used interchangeably to describe systems ranging from simple Proportional-Integral-Derivative (PID) feedback loops to generative agents. This imprecision prevents funding bodies and facility directors from accurately assessing technical risk or defining liability. For instance, while a simple script lacks the agency to distinguish between a sample and a beamstop, a generative agent actively navigates physical space. Without a standardised taxonomy to quantify this agency, facilities cannot define the Operational Design Domains (ODD) required to insure and license autonomous experiments.

Existing taxonomies for autonomous experimentation are predominantly predicated on an owner-operator model, typically used in electron microscopy~\cite{kalinin2021automated}. In these scenarios, a principal investigator retains instrumental sovereignty and can afford extended downtime to train a reinforcement learning agent \textit{in-situ}. Under the constraints of a transient user model, where beamtime is limited to 48–96 hours, dedicating significant periods to on-policy training is operationally impractical. This operational constraint renders generic self-driving laboratory taxonomies incompatible with the large-scale facility context. To address this constraint, User Facilities require a zero-shot deployment capability. This is most effectively realised through Sim-to-Real transfer, where agents are pre-trained on physics-based Digital Twins~\cite{kalidindi2022digital} to ensure immediate functionality upon deployment.

Finally, large-scale facilities face a computational constraint that distinguishes them from direct imaging techniques: the Inference Barrier. Decisions in X-ray tomography and diffraction cannot be derived from raw detector data such as sinograms or diffractograms; they require real-time inversion into physical space to be scientifically valid. This necessity creates a critical latency gap, forcing agents to operate on physics-based Digital Twins rather than raw sensor streams. To systematise the capabilities required to navigate this constraint, we propose the Benchmarking Autonomy in Scientific Experiments (BASE) scale. This hierarchical taxonomy not only quantifies experimental intelligence but defines the operational boundaries required for zero-shot autonomy and secure federated learning. This work serves as a theoretical blueprint and architectural specification for the next generation of autonomous facility infrastructure.

\section*{The BASE Taxonomy of Agency}

\subsection*{The Definition of Agency vs. Automation}

The BASE scale benchmarks the transfer of cognitive load from the human operator to the experimental system. It does not measure mechanical complexity or data throughput. A critical distinction is drawn between automation and autonomy. Automation refers to the computerised execution of predefined tasks with high precision and speed. Autonomy refers to the ability of a system to make decisions regarding which tasks to execute based on environmental feedback. A robotic sample changer operating at 100 Hz represents a high degree of automation but possesses zero autonomy if the execution sequence is rigidly scripted.

Recent work has demonstrated the utility of Large Language Models in automating accelerator operations. Hellert et al. successfully deployed an agentic system at the Advanced Light Source capable of translating natural language into executable control scripts for machine physics tasks~\cite{hellert2024application}. This development validates the supervisory model for machine control, where an agent orchestrates subsystems to optimise beam parameters based on scalar process variables.

However, a critical gap remains in applying this agency to the scientific experiment itself. Unlike accelerator physics, where decisions are driven by scalar variables such as magnet current or vacuum pressure, experimental autonomy requires decisions based on complex, high-dimensional data streams. In X-ray tomography or diffraction, the decision manifold is defined not by raw sensor output, but by reconstructed physical properties. Consequently, the agent must overcome the computational latency of inference to close the feedback loop. The BASE scale explicitly defines the hierarchy of agency required to bridge this gap between scalar machine control and multidimensional experimental discovery.

\begin{figure}[ht]
\centering
\includegraphics[width=\linewidth]{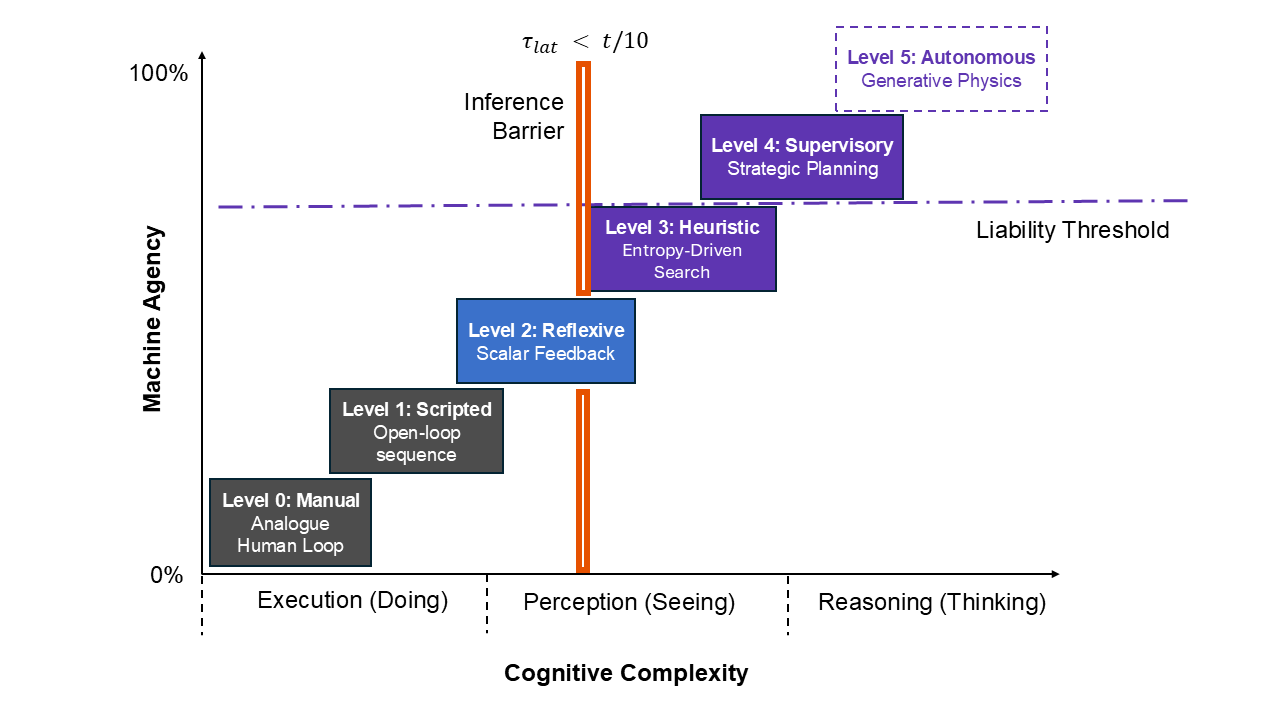} 
\caption{The Benchmarking Autonomy in Scientific Experiments (BASE) Scale. 
A hierarchical taxonomy of experimental agency adapted for Large-Scale Facilities. The scale progresses from Level 0 (Manual) to Level 5 (Autonomous). 
The critical operational discontinuity occurs at the Inference Barrier (Orange), where the system must invert raw data within the latency budget ($\tau_{lat} < t/10$) to enable Level 3 (Heuristic) control. 
At this level, agents use physics-based priors (Entropy-Driven Search) to target information-rich features. 
The Liability Threshold (Purple) marks the transition from human-validated safety to algorithmic safety, requiring a rigorous Safety Case before Level 4 (Supervisory) deployment.}
\label{fig:staircase}
\end{figure}

\begin{table}[ht]
\centering
\small 
\renewcommand{\arraystretch}{1.3} 
\begin{tabular}{@{} p{0.08\linewidth} p{0.15\linewidth} p{0.18\linewidth} p{0.20\linewidth} >{\raggedright\arraybackslash}p{0.32\linewidth} @{}}
\toprule
\textbf{Level} & \textbf{Name} & \textbf{Execution} \newline \textit{(Who moves?)} & \textbf{Monitoring} \newline \textit{(Who interprets?)} & \textbf{Operational Capability} \\
\midrule
\textbf{L0} & Manual & Human & Human & \textbf{Reactive:} User manually performs alignment, acquisition, and data management. \\
\addlinespace[6pt]

\textbf{L1} & Scripted & System & Static Logic & \textbf{Deterministic Execution:} System executes a static instruction set (e.g. grid scan). Continues regardless of sample evolution or failure. \\
\addlinespace[6pt]

\textbf{L2} & Reflexive & System & Scalar Feedback & \textbf{Assisted Execution:} System optimises parameters via closed-loop feedback on scalar variables (e.g. auto-focus, beam-alignment). \\
\addlinespace[6pt]

\textbf{L3} & Heuristic & System & Inference Engine & \textbf{Conditional Autonomy:} System identifies and targets specific morphological features within a defined ODD using a physics-based Digital Twin. \\
\addlinespace[6pt]

\textbf{L4} & Supervisory & System & Strategic Agent & \textbf{Campaign Management:} System formulates experimental strategy across multiple samples. Human remains in-the-loop for high-level liability. \\
\addlinespace[6pt]

\textbf{L5} & Autonomous & System & System \newline (Generative) & \textbf{Closed-Loop Discovery:} System synthesises hypotheses and updates physical models via independent agency, operating without human intervention. \\ \\
\bottomrule
\end{tabular}
\caption{The BASE Scale Definition Matrix. The transition from Automation (L1--L2) to Autonomy (L3--L5) is defined by the shift from scalar feedback to model-based perception. L3 represents the crossing of the Inference Barrier.}
\label{tab:base_matrix}
\end{table}

\subsection*{Levels 0--2: Deterministic and Reactive Control}

The lower tiers of the BASE scale are defined by the optimisation of task execution. In this regime, the system focuses on the efficiency of a predefined procedure but lacks the agency to evaluate the scientific utility of the data as it is acquired.

\subsubsection*{Level 0: Manual Operation}
Level 0 represents a baseline manual environment where the human operator executes the Observe-Orient-Decide-Act (OODA) loop in its entirety. Experimental throughput is strictly constrained by human reaction time and the inherent limits of manual coordination. This introduces a fatigue-induced bottleneck, where data acquisition remains reactive. Critical decision-making tends to degrade during extended operational shifts, as the temporal resolution of human attention is insufficient to maintain the continuous vigilance required to capture transient physical events.

\subsubsection*{Level 1: Scripted Automation} 
Level 1 represents the dominant operational mode for high-throughput user facilities. The system executes rigid, predefined command sequences using frameworks such as GDA~\cite{gibbons2011gda} or Bluesky~\cite{allan2019bluesky}. While this maximises acquisition frequency by eliminating human latency, it introduces an open-loop vulnerability where the system operates without semantic awareness of the data it acquires. Consequently, if a sample degrades or beam conditions fluctuate, the script continues execution regardless, resulting in the accumulation of high-volume but scientifically void datasets.

The data volume bottleneck arises as a direct consequence of this throughput-centric approach. By prioritising raw data velocity over perception, Level 1 automation industrialises the collection of redundancy rather than the extraction of insight. As Drnec and Lyonnard observe, such workflows exacerbate the systemic crisis in Reliability, Representativeness, and Reproducibility~\cite{drnec2025battery}. Economically, these inefficiencies impose a significant burden on facility infrastructure, as substantial resources are consumed storing high-volume, low-entropy datasets generated by rigid, uniform grid-sampling~\cite{lu2025autonomousbatteryresearchprinciples}. Under this paradigm, the system effectively optimises for the acquisition of data lacking scientific information density.

\subsubsection*{Level 2: Reflexive Assistance} 
Reflexive assistance introduces closed-loop control based on real-time sensor feedback. At this level, the system uses signal processing to optimise execution parameters without a semantic understanding of the scientific object. Control logic relies on scalar signals; typical applications include maximising integrated intensity for auto-focus routines or calculating the centre of mass for beam-sample alignment. While these systems are responsive, they remain confined to optimising the measurement process rather than discovering the underlying physics.

The distinction at this level lies between signal and semantics. The system is capable of centring a rotation axis based on a sinogram centre of mass calculation, but it lacks the world model to identify the sample being rotated. Importantly, these operations use raw data streams with millisecond-scale latency. Decisions rely on scalar process variables or raw detector outputs, avoiding the computational overhead of full tomographic reconstruction or spectral inversion.

\subsection*{Level 3: The Heuristic Threshold}

The transition from Level 2 to Level 3 represents a fundamental discontinuity in the experimental feedback loop defined here as the Inference Barrier. Unlike autonomous automotive systems where optical sensors capture immediately recognisable features, such as a stop sign visible in a raw image, detectors acquire raw projections like sinograms or diffractograms. These data are unintelligible in their raw form.

\subsubsection*{Formalising the Latency Budget}

The transition to Level 3 autonomy presents a control theory problem defined by the total latency budget. We model the experimental feedback loop as a sum of sequential delays. To exercise heuristic control over a physical process with a characteristic timescale $t_{\text{dynamics}}$, the total loop latency $\tau_{\text{loop}}$ must satisfy the controllability criterion:

\begin{equation} \tau_{\text{loop}} = \tau_{\text{readout}} + \tau_{\text{transport}} + (\tau_{\text{reduce}} + \tau_{\text{inference}}) + \tau_{\text{actuation}} \leq \frac{t_{\text{dynamics}}}{\kappa} \label{eq:latency_budget} \end{equation}

Here $\kappa$ denotes the control safety factor, typically set to $\kappa \geq 10$ to prevent temporal aliasing. To be precise, these terms are governed by fundamental hardware and physical constraints. The detector readout latency $\tau_{\text{readout}}$ accounts for the sensor deadtime and internal processing before transmission begins. The data transport latency $\tau_{\text{transport}} \approx \mathcal{D}_{\text{frame}}/\mathcal{B}_{\text{net}}$ is defined by the ratio of the detector frame size $\mathcal{D}_{\text{frame}}$ to the network bandwidth $\mathcal{B}_{\text{net}}$. For a 100 MB frame on a standard 10 GbE link, this introduces an irreducible delay of approximately 80 ms which often violates the budget before computation begins. The computational overhead is the sum of $\tau_{\text{reduce}}$ and $\tau_{\text{inference}}$, representing the time required to invert raw data into physical quantities such as Fourier transforms or reconstructions and the neural network forward-pass time respectively. This constitutes the primary target for edge-computing acceleration. Actuation inertia $\tau_{\text{actuation}}$ accounts for the electromechanical settling time of the beamline hardware. Finally, the process dynamics $t_{\text{dynamics}} \approx \delta x / v_{\text{evol}}$ are defined by the ratio of the required spatial resolution $\delta x$ to the velocity of the physical evolution $v_{\text{evol}}$. For example, capturing a crack propagating at 1 $\mu$m/s with 100 nm resolution yields a characteristic timescale $t_{\text{dynamics}} \approx 100$ ms. To satisfy the safety factor $\kappa=10$, this requires a loop frequency exceeding 100 Hz, demanding a total latency $\tau_{\text{loop}} < 10$ ms. Violating this inequality where $\tau_{\text{loop}} > t_{\text{dynamics}}/\kappa$ forces a phase transition from active Level 3 control to passive Level 1 monitoring.

The necessity of closing the feedback loop introduces a computational latency gap, reframing facility autonomy from a robotics challenge into a high-performance computing challenge. While millisecond-latency inversion is technically feasible for 2D diffraction or imaging, full 3D tomographic reconstruction at comparable frequencies currently approaches the theoretical limits of edge infrastructure. Therefore, effective Level 3 feasibility frequently depends on the utilisation of proxy signals. In this regime, the agent operates on lower-dimensional surrogates, such as sparse sinograms or orthogonal 2D projections to infer 3D interest without necessitating full volumetric reconstruction within the control loop ($\tau_{loop}$). By decoupling decision latency from reconstruction latency, the system maintains heuristic control even when the full inversion time violates the strict observability criterion.

\subsubsection*{Conditional Autonomy and Physics-Aware Priors} 
Once the inference barrier is crossed, Level 3 is defined as conditional autonomy. At this stage, the system actively modifies the experimental strategy to resolve specific scientific features defined by the user. Unlike the scalar feedback of Level 2, Level 3 is physics-aware. It uses semantic priors to identify features that are statistically rare but scientifically critical. Importantly, the Digital Twin is not intended to replace the experiment but to benchmark the standard operating state. The heuristic agent detects scientific value precisely when the experimental observation diverges from the Digital Twin's prediction, flagging a high-entropy anomaly.

\begin{figure}[ht]
\centering
\includegraphics[scale = 0.6]{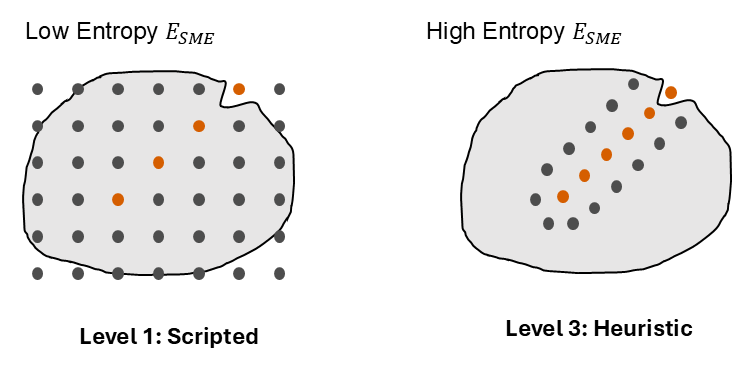}
\caption{The Information Efficiency Gap. (a) Level 1 (Scripted Automation): A traditional dense raster scan applies equal incident flux to all spatial voxels. In deterministic failure modes, this wastes $>90\%$ of beamtime measuring the stable, elastic bulk. (b) Level 3 (Semantic Autonomy): An active heuristic agent uses a physics-based prior (e.g., surface kurtosis) to drive an adaptive scan. The agent autonomously clusters measurement points around the information-rich stress concentrators (notches), maximising Entropy-Scaled Measurement Efficiency ($E_{\text{SME}}$).}
\label{fig:efficiency_gap}
\end{figure}

For example, recent work on high-nickel cathodes demonstrates that particle failure is driven by stress concentrators defined by geometric kurtosis~\cite{LeHoux2026}. A Level 3 agent could use this prior to actively optimise the scan trajectory, resolving this specific topography with high fidelity rather than rastering the entire particle volume. This approach maximises the information density of the experiment by focusing flux solely on the features governing the failure mechanism, as seen in figure \ref{fig:efficiency_gap}.

Autonomy at this level remains bounded. The agent operates strictly within a specific Operational Design Domain (ODD) defined as a set of environmental, physical, and safety parameters. If the entropy metric of the incoming data stream falls outside model confidence bounds, or if the inversion latency violates the observability criterion (where $\tau_{\text{loop}} > t_{\text{dynamics}}/\kappa$), the system triggers an immediate reversion to human control.

\subsection*{Levels 4--5: Strategic Agency}

The upper tiers of the BASE scale mark the transition from semantic perception to strategic agency. This shift reorients the focus from quantifying data volume to defining the authority over the experimental campaign.

\subsubsection*{Level 4: Supervisory Autonomy} 
Level 4 designates supervisory autonomy, frequently described as the Guardian Angel or concierge model~\cite{kalinin2024human}. At this stage, the agent manages the campaign strategy over extended durations, prioritising sample batches based on stability or data quality. The operational logic follows a human in-the-loop paradigm where the system proposes strategic adjustments while the human operator retains veto power over significant deviations. Consequently, the human maintains ultimate operational authority for liability purposes.

The transient user model inherent to large-scale facilities necessitates a distinct approach to agent training. Unlike owner-operator laboratories, where a principal investigator may afford extended downtime to train a reinforcement learning agent \textit{in-situ}~\cite{kalinin2021automated}, a facility user with limited beamtime requires immediate functionality. Consequently, the Level 4 agent acts as a concierge, arriving pre-trained on generalised physical models to enable zero-shot deployment. This capability distinguishes facility-grade autonomy from specialist agents requiring extensive localised optimisation.

However, deploying active agents introduces a significant shift in liability. While a Level 1 script is deterministic, a Level 4 agent operates probabilistically. If an agent drives a motor beyond safe limits or damages a unique sample environment, the liability extends beyond the operator to the validation framework that certified the agent's logic. This necessitates the development of a rigorous safety case for every autonomous system deployed on the beamline.

\subsubsection*{Level 5: Full Autonomy} 
Level 5 designates full autonomy, synonymous with the self-driving laboratory. In this regime, the system synthesises the hypothesis, executes the experiment, and updates the physical model without human intervention. This model of unsupervised discovery was recently demonstrated by the A-Lab platform~\cite{szymanski2023autonomous}, where the scientific loop operates entirely independent of human supervision.

While this model represents the aspirational standard for owner-operator chemistry or microscopy laboratories, it poses an unacceptable operational risk for multi-user national facilities. The significant cost of failure at an experimental endstation, combined with the heterogeneity of user experiments, renders unconstrained curiosity-driven discovery impractical. The BASE scale therefore prioritises the heuristic efficiency of Level 3 and the supervisory safety of Level 4 over the complete independence of Level 5. Fundamentally, the objective of facility autonomy is not to replace the scientist but to elevate the researcher from machine operator to experimental architect.

\section*{Information Gain as a Control Logic}

Recalling the degradation mechanism in high-nickel cathodes~\cite{LeHoux2026}, failure is not a stochastic bulk phenomenon but is deterministically driven by rare geometric features. Specifically, surface notches act as stress concentrators, shown to generate localised stresses of approximately 85 MPa, exceeding the material's 39 MPa yield strength by a factor of two. Importantly, this plasticity is spatially confined: while the notch tip yields, the bulk of the particle accommodates the strain elastically.

From an information theory perspective, a Level 1 grid scan measuring the stable bulk is information-inefficient. It wastes over 99\% of the available flux confirming the known prior that the bulk is elastic. In contrast, a Level 3 semantic agent maximises Entropy-Scaled Measurement Efficiency ($E_{\text{SME}}$), as mathematically formalised in the Heuristic Operando framework~\cite{lu2025autonomousbatteryresearchprinciples}. By using physics-based priors to actively target regions of high variance, the agent maximises the reduction of epistemic uncertainty regarding specific features (such as dendrites or notches) while ignoring regions dominated by aleatoric noise.

To formalise this heuristic, we define the control variable as the spectral kurtosis of the diffraction peak. For a measured intensity distribution $I(q)$ with mean $\mu$ and standard deviation $\sigma$, the excess kurtosis $g_2$ is calculated live on the FPGA:

\begin{equation} g_2 = \frac{\frac{1}{n} \sum_{i=1}^{n} (I_i - \mu)^4}{\left( \frac{1}{n} \sum_{i=1}^{n} (I_i - \mu)^2 \right)^2} - 3 \label{eq:kurtosis} \end{equation}

Physically, the stable bulk material follows a Gaussian distribution where $g_2 \approx 0$. In contrast, the strain fields surrounding a stress concentrator cause non-Gaussian peak broadening, manifesting as a heavy-tailed outlier where $g_2 \gg 0$.

Consequently, this Level 3 agent does not require a hard-coded asymptotic limit. Instead, it dynamically calculates a decision boundary $K_{\text{thresh}} = \mu_{\text{bulk}} + 3\sigma_{\text{bulk}}$ based on the initial stream of spherical particles. The agent actively hunts for high-kurtosis signatures, rejecting spherical, low-kurtosis particles to focus acquisition solely on the failure sites. This transition defines the boundary between Level 1 and Level 3: Level 1 maximises coverage in terabytes per hour, whereas Level 3 maximises insight in entropy per hour. By shifting the control logic from a spatial grid to an information gradient, we enable heuristic operando experiments that are impossible to perform manually.

\subsection*{Mathematical Definition of Heuristic Autonomy}
To enable rigorous benchmarking, we formalise the autonomous experimental task as a constrained optimisation problem. Replacing rigid path-planning with dynamic decision-making, a Level 3 agent executes a control policy $\pi$ that maps a stream of observations $\mathbf{o}_t$ to experimental actions $\mathbf{a}_t$. The objective of the agent is to maximise the Entropy-Scaled Measurement Efficiency ($E_{\text{SME}}$) over the campaign horizon $T$. However, the cost of an experiment is rarely univariate. It comprises a vector of competing resources: wall-clock time, cumulative radiation dose, inference compute cycles, and data storage volume. To make the optimisation tractable, we define a resource cost vector $\mathbf{C}(\mathbf{a})$. We therefore propose the Heuristic Policy Function:

\begin{equation}
\pi^* = \operatorname*{argmax}_{\pi} \mathbb{E}_{\mathbf{D} \sim P_{\text{sim}}} \left[ \frac{I(M; \mathbf{D}_\pi)}{\mathbf{w}^\intercal \mathbf{C}(\mathbf{a})} \right]
\label{eq:governing_eq}
\end{equation}

\noindent subject to the operational constraints:

\begin{equation}
\begin{cases} 
\mathbf{s}_t \in \mathcal{O}_{\text{ODD}} & \text{(Safety Constraint)} \\ 
\tau_{\text{loop}} \leq \frac{t_{\text{dynamics}}}{\kappa} & \text{(Observability Constraint)} 
\end{cases}
\end{equation}

Here, the numerator $I(M; \mathbf{D}_\pi)$ quantifies the mutual information gained regarding the scientific hypothesis space $M$. The expectation $\mathbb{E}$ is evaluated over the distribution of predicted future observations $\mathbf{D}_\pi$. This dependency explicitly enforces the architectural requirement for a Level 3 Digital Twin ($P_{\text{sim}}$). Without a generative model to forecast the distribution of unmeasured data, the agent cannot compute this expectation to plan its trajectory.

The denominator represents the scalarised cost, where $\mathbf{C}(\mathbf{a}) = [t_{\text{acq}}, \mathcal{D}_{\text{dose}}, V_{\text{data}}, \mathcal{C}_{\text{inf}}]^\intercal$ denotes the vector of resource consumption (time, dose, storage, and compute cycles, respectively). The weight vector $\mathbf{w}$ functions as a user-defined hyperparameter that dictates the operational mode. For instance, setting $\mathbf{w} = [0.1, 10, 0, 0]$ establishes a low-dose mode for beam-sensitive samples, whereas $\mathbf{w} = [10, 0, 0, 0]$ imposes a high-throughput mode. The constraints explicitly formalise the boundary conditions defined by the taxonomy: the physical state of the system $\mathbf{s}_t$ must remain within the valid Operational Design Domain $\mathcal{O}_{\text{ODD}}$ (Level 4 requirement), and the total loop latency must adhere to the Nyquist-Shannon observability criterion derived in Equation \ref{eq:latency_budget} (Level 3 requirement). This formalism provides a mathematically precise definition of autonomy: a system is effectively autonomous only if it can solve Equation \ref{eq:governing_eq} without human intervention while strictly satisfying both boundary conditions.

\section*{Operational Design Domains and Liability}

Artificial intelligence in scientific contexts is often viewed as a binary capability. However, this perspective can oversimplify the safety requirements of active control systems. To ensure operational integrity, we adopt the automotive concept of Operational Design Domains (ODD). Within this framework, a Level 3 agent functions autonomously only within specified environmental parameters, such as defined sample taxonomies, energy ranges, or kinematic limits. If the operational state exceeds this ODD, the system is designed to revert to human control. Accordingly, defining these boundaries becomes a primary requirement for deploying autonomous agents at national facilities.

\subsection*{The Safety Case for Active Agents} 
We differentiate between passive agents, used for analysis or data reduction, and active agents authorised for beam steering or hardware control. Passive agents entail minimal operational risk. In contrast, an active agent executing a curiosity-driven policy possesses the physical capacity to induce kinematic collisions, damage sample environments, or cause catastrophic beam loss, potentially precipitating thermal shock to insertion devices. Such failures incur substantial costs in facility downtime and hardware repairs. Consequently, a Level 4 agent cannot be deployed on a high-value experimental endstation without a rigorous safety case. The fail-fast methodologies common in commercial software are incompatible with the high-consequence environment of a particle accelerator. Furthermore, purely data-driven agents risk generating confident hallucinations when operating outside their training distribution. Therefore, validation protocols must strictly penalise agents that fail to trigger a safe fallback mode when the internal model diverges from experimental reality.

\subsection*{Sim-to-Real Certification}
To mitigate this risk, we propose a mandatory validation protocol. Before an agent is granted write-access to instrument control systems, it must pass a standardised competency assessment within a Virtual Beamline. This digital twin simulates the physics, hardware kinematics, and failure modes of the real instrument. To achieve certification, the agent must demonstrate the ability to maximise Entropy-Scaled Measurement Efficiency ($E_{\text{SME}}$) without violating hard safety constraints, such as motor collisions or beam loss events~\cite{lu2025autonomousbatteryresearchprinciples}.

Operationally, this framework decouples safety verification from the manual oversight of the beamline scientist. Instead, users are required to submit their control policies to the validation pipeline prior to their allocated beamtime. Access to physical hardware is subsequently granted on a conditional basis, contingent upon the successful completion of these simulation benchmarks.

\subsection*{Operational Assurance}

As experimental control shifts from Level 1 execution to Level 4 supervision, the locus of operational risk evolves. In manual experiments, safety relies on human vigilance; in autonomous experiments, safety relies on the robustness of the Operational Design Domain (ODD). To define operational assurance within this high-complexity environment, we explicitly distinguish between scientific intent and kinematic safety. Given the non-deterministic nature of experimental research, a facility cannot guarantee the predictive accuracy of a digital twin regarding complex physical phenomena. Instead, the operational mandate must focus on the provision of a deterministic safety envelope.

Under this framework, liability is strictly decoupled from algorithmic performance. The facility assumes responsibility for maintaining hardware-level interlocks—defined as the kinematic hard-limits of the ODD—that physically prevent unsafe actuation regardless of the control input. Consequently, if an agent attempts an unsafe action, the liability lies with the facility’s safety infrastructure for failing to reject the command, rather than with the user’s algorithmic logic. This shift encourages facilities to invest in rigorous, hardware-based containment rather than restricting the complexity of user-deployed agents.

Reflecting this architectural separation, we propose a new framework for algorithmic governance. While the hardware ensures safety, the Digital Twin ensures transparency. If an agent successfully passes certification but fails during physical operation, the failure mode indicates a critical deficiency in the simulation fidelity or the Sim-to-Real transfer function. Accordingly, the primary engineering mandate for the facility expands from maintaining physical hardware to curating a high-fidelity Digital Twin~\cite{kalidindi2022digital}. By treating the simulation environment as critical infrastructure, facilities can cultivate a secure ecosystem for third-party agents, effectively resolving the governance bottleneck that currently impedes the widespread adoption of autonomous experimentation.

\subsection*{Formalising the Verification Gap}
To strictly define the limits of operational assurance, the validation protocol requires a quantitative metric for the fidelity of the Digital Twin. While the theoretical divergence between the simulation and reality is canonically described by the Kullback-Leibler divergence ($D_{\text{KL}}$), this metric is operationally intractable in real-time because the true physical distribution $P_{\text{real}}$ is unknown. Computing $D_{\text{KL}}$ requires an ensemble of identical actions to empirically estimate $P_{\text{real}}$, a requirement that is impossible to satisfy during a unique, transient experimental failure. Consequently, we define the Sim-to-Real Gap $\Delta_{\text{gap}}$ using the Negative Log-Likelihood (NLL) of the single observed real-world state $\mathbf{s}'_{\text{obs}}$ given the simulator's prediction. This formulation effectively measures the surprisal or information content of the failure event relative to the model:

\begin{equation} \Delta_{\text{gap}}(\mathbf{s}t, \mathbf{a}t, \mathbf{s}'{\text{obs}}) = - \ln P{\text{sim}}(\mathbf{s}'_{\text{obs}} | \mathbf{s}_t, \mathbf{a}_t) \label{eq:sim_to_real} \end{equation}

By quantifying the verification gap, we establish an algorithmic basis for liability. We define a certified validity bound $\epsilon_{\text{valid}}$, calibrated during the Digital Twin's commissioning phase using offline verification benchmarks (where the computation of $D_{\text{KL}}$ is feasible). If an active agent subsequently causes hardware damage at state $\mathbf{s}_t$ resulting in an outcome $\mathbf{s}'_{\text{obs}}$, liability $\mathcal{L}$ is assigned according to the following boolean condition:

\begin{equation}
\mathcal{L} = 
\begin{cases} 
\text{Facility}, & \text{if } \Delta_{\text{gap}} > \epsilon_{\text{valid}} \quad \text{(Model Failure: Outcome physically unpredicted)} \\ 
\text{User}, & \text{if } \Delta_{\text{gap}} \leq \epsilon_{\text{valid}} \quad \text{(Policy Failure: Prediction ignored)}
\end{cases}
\label{eq:liability_logic}
\end{equation}

Under this formalism, the division of responsibility is explicit. Provided the simulator assigns a reasonable probability to the adverse outcome, characterised by a low NLL, the user is liable for any unsafe control policy $\pi$ that navigates the system into that state. For instance, this applies if the policy ignores a collision warning correctly predicted by the simulation. Conversely, if the simulator assigns a negligible probability to the event, yielding a high NLL and effectively presenting a misleading safe zone, the facility assumes liability for the model discrepancy. Operationally, this mandates that the facility maintain a secure data logger to record the trajectory tuple $(\mathbf{s}_t, \mathbf{a}_t, P_{\text{sim}}, \mathbf{s}'_{\text{obs}})$ for post-incident forensic analysis.

\subsection*{Handling the Exploration Paradox}

A fundamental tension exists between the requirement for rigid Operational Design Domains and the exploratory nature of scientific discovery. While a Level 4 agent relies on a bounded world model to ensure safety, frontier research frequently aims to probe beyond these established boundaries to discover unknown physics, denoted as $m_{\emptyset}$. However, purely entropy-driven agents face a critical vulnerability in which the system maximises epistemic uncertainty by pursuing high-entropy stochastic noise, such as detector glitches or beam dumps, rather than genuine physical phenomena.

To resolve this paradox, we impose a hierarchical validation logic that differentiates between instrument failure (aleatoric uncertainty) and scientific novelty (epistemic uncertainty). We define the discovery potential, $D(x)$, not merely as a function of signal entropy $H(x)$, but as a conditional property of the instrument state vector, $\mathbf{s}_{t}$.

\begin{equation}
D(x) = H(x) \cdot \mathbb{I}(\mathbf{s}_{t} \in \Omega_{\text{nominal}})
\end{equation}

Here, $\mathbb{I}$ represents an indicator function that returns 1 only if the instrument state vector, comprising incident flux $I_0$, detector saturation levels, and motor following errors, resides strictly within the nominal Operational Design Domain, $\Omega_{\text{nominal}}$. High-entropy events detected while $\mathbf{s}_{t}$ deviates from this domain are immediately classified as instrumental artefacts.

Operationally, this logic enforces a non-vanishing prior on instrument failure. When an agent encounters a high-entropy anomaly, it must trigger an immediate calibration check to satisfy the indicator function before the observation can be accepted as a valid input for the scientific model. Consequently, we propose a staged deployment strategy. Level 4 autonomy is initially restricted to deterministic operating modes, such as parametric mapping. Experiments probing undefined failure modes must remain at Level 3, retaining a human-in-the-loop validation step until the generative world model achieves sufficient generalisation to distinguish physical anomalies from hardware variance.

\section*{Infrastructure Requirements for the Semantic Web}

Existing facility data infrastructures predominantly rely on passive archival formats such as HDF5 or NeXus. While these standards excel at high-throughput storage, they remain semantically opaque to autonomous agents. An autonomous policy cannot operate effectively if it is required to parse unstructured string headers to infer critical state parameters, such as beam energy or sample-detector distance. Such data opacity creates a fundamental misalignment between Level 1 data structures and the semantic perception required for Level 3 autonomy. To resolve this disconnect, the community must transition from static data catalogues to dynamic knowledge graphs that provide intrinsic, machine-readable context.

To enable zero-shot autonomy for transient users, facility hardware must evolve to be self-describing. We propose a transition toward intelligent data formats where the complete instrument definition, including geometry, kinematic constraints, and safety limits, is either embedded directly within the data stream or broadcast via a standardised ontology. Such a semantic interface is a prerequisite for a plug-and-play operational model, allowing an agent to arrive at a new endstation and instantly query capabilities like maximum energy or motor ranges without manual configuration. This architectural shift eliminates the need for visiting agents to dedicate valuable beamtime to learning unique motor nomenclatures, thereby reducing the integration overhead from days to minutes.

Beyond hardware definitions, a Level 3 agent requires access to scientific priors to execute informed decision-making. For instance, complex domain knowledge, such as the geometric definition of a stress concentrator described in recent work, cannot be rigidly hard-coded into every individual experimental script~\cite{LeHoux2026}. Instead, these priors must reside in a federated semantic atlas that the agent can query in real-time. Adopting this federated learning architecture also provides a robust solution to the challenge of industrial data sovereignty. By inverting the standard data flow, the framework allows the model to travel to the data rather than requiring the data to travel to the model~\cite{lu2025autonomousbatteryresearchprinciples}. This architecture enables a facility to host an agent that learns from proprietary industrial streams, such as those from gigafactories, without sensitive information ever leaving the secure local premise. Such an approach effectively resolves the privacy bottleneck that currently stalls the deployment of industrial AI at user facilities.

Facilities must reorient their funding strategies, moving away from isolated AI pilot projects that frequently stagnate within code repositories. Instead, resources should be directed toward the development of autonomous infrastructure. Key priorities must include the fundamental engineering tasks of constructing semantic middleware, standardising ontologies, and certifying the digital twins that serve as the operational environment for these agents. Ultimately, the trajectory toward the self-driving laboratory will not be defined by the sophistication of neural networks, but by the rigour of the underlying data standards.

\section*{A Roadmap for the Self-Driving Beamline}

The transition from automated data collection to autonomous discovery necessitates a shift in discourse, moving from vague aspirations of AI for Science to concrete engineering specifications. Through the adoption of the BASE scale, the community gains the ability to distinguish between a smart script at Level 1 and a semantic agent at Level 3 with rigorous technical precision. Such a standardised vocabulary empowers facility directors and funding bodies to draft precise requirements, enabling the commissioning of Level 3-capable endstations with explicitly defined Operational Design Domains, rather than soliciting undefined autonomous beamlines.

A central challenge in the deployment of autonomous facilities lies in defining an Operational Design Domain for exploratory research, particularly when the underlying physics is inherently unknown. However, the hierarchical validation logic established in the previous section provides a rigorous mechanism to navigate this uncertainty. By treating the unknown as a distinct, mathematically tractable hypothesis ($m_{\emptyset}$) strictly conditional upon the instrument state vector $\mathbf{s}_{inst}$, facilities can move beyond the false dichotomy of safe automation versus risky autonomy. This architectural approach permits the commissioning of Level 3 agents explicitly designed to optimise for Entropy-Scaled Measurement Efficiency, enabling a mode of operation where the machine handles the search for statistical outliers while the human architect retains authority over the scientific interpretation.

\subsection*{Benchmarking the State-of-the-Art}

To demonstrate the utility of the BASE scale as a comparative tool, we apply the taxonomy to classify several prominent autonomous frameworks currently deployed at large-scale facilities. By mapping these systems onto the scale, we aim to resolve the ambiguity between sophisticated parameter optimisation and true semantic autonomy.

\begin{table}[ht]
\centering
\small 
\renewcommand{\arraystretch}{1.3} 
\begin{tabular}{@{} p{0.18\linewidth} p{0.20\linewidth} p{0.15\linewidth} >{\raggedright\arraybackslash}p{0.42\linewidth} @{}}
\toprule
\textbf{System} & \textbf{Reference} & \textbf{BASE Level} & \textbf{Classification Rationale} \\
\midrule
\textbf{Adaptive XRD} & Szymanski et al.~\cite{szymanski2023adaptively} & \textbf{Level 2} \newline (Reflexive) & \textbf{Reactive Optimisation:} The system uses model uncertainty to resample phase boundaries. It optimises instantaneous sampling density based on current feedback but does not utilise a temporal digital twin to predict future states. \\
\addlinespace[6pt]

\textbf{Sparse Scanning} & Yager et al.~\cite{yager2023autonomous} & \textbf{Level 2} \newline (Reflexive) & \textbf{Efficiency Enhancement:} The agent optimises spatial sampling to maximise signal-to-noise. This is a reflexive improvement of \textit{how} to scan, rather than a semantic decision of \textit{what} or \textit{when} to scan. \\
\addlinespace[6pt]

\textbf{ANDiE} & McDannald et al.~\cite{mcdannald2022fly} & \textbf{Level 3} \newline (Heuristic) & \textbf{Physics-Aware Hunting:} The agent utilises specific physics priors (Weiss/Ising models) to actively hunt for a magnetic phase transition. It constrains the search path to avoid hysteresis, demonstrating semantic awareness. \\
\addlinespace[6pt]

\textbf{Edge-to-Exascale} & Yin et al.~\cite{yin2024integrated} & \textbf{Level 3} \newline (Heuristic) & \textbf{Predictive Control:} The system employs a Temporal Fusion Transformer to forecast the evolution of neutron scattering patterns. By acting on a predicted future state rather than current feedback, it achieves heuristic control. \\
\addlinespace[6pt]

\textbf{Gaussian Process MAPs} & Maffettone et al.~\cite{maffettone2023self} & \textbf{Level 2+} \newline (Reflexive) & \textbf{Combinatorial Search:} While highly capable, these agents typically minimise posterior uncertainty over a static parameter space. They lack the specific anomaly-detection logic required for Level 3 event capture. \\
\bottomrule
\end{tabular}
\caption{\textbf{Classification of State-of-the-Art Systems.} Existing autonomous frameworks are differentiated by the BASE scale. Level 2 systems focus on optimising data fidelity or coverage (Reflexive), while Level 3 systems utilise physics-based priors to actively hunt for specific scientific features (Heuristic).}
\label{tab:sota_benchmarking}
\end{table}

The classification presented in Table \ref{tab:sota_benchmarking} indicates that the majority of currently deployed autonomous workflows operate at Level 2 (Reflexive). While these systems excel at optimising data quality, typically by maximising signal-to-noise ratios or efficiently mapping phase boundaries via uncertainty quantification, they remain fundamentally reactive. In contrast, true Level 3 (Heuristic) autonomy, as demonstrated by the ANDiE and SNS Edge-to-Exascale frameworks, necessitates the integration of physics-based priors (e.g., Ising models or temporal transformers). These priors enable the agent to predict and actively target specific scientific phenomena, transcending simple scalar feedback loops.

Furthermore, the BASE scale clarifies that autonomy functions as a hierarchical dependency stack rather than a simple binary capability. Consequently, semantic perception at Level 3 remains unattainable without first resolving the inference barrier inherent to Level 2. By structuring these dependencies, the taxonomy exposes critical infrastructure gaps, most notably the absence of real-time digital twins and semantic data layers, that currently impede the community's transition from high-throughput to high-insight operations.

While Level 5 fully autonomous discovery remains a theoretical horizon, the immediate strategic value for national facilities lies in mastering heuristic search at Level 3 and supervisory agency at Level 4. These tiers deliver the order-of-magnitude efficiency gains required to resolve deterministic failure modes without incurring the unacceptable operational risks associated with fully unsupervised experimentation. Furthermore, the standardisation of the BASE scale creates a commercial incentive for instrument vendors to integrate Guardian Angel safety modules directly into detector hardware, effectively shifting the safety burden from user code to the infrastructure itself. Ultimately, the objective is not to replace the scientist, but to safely elevate their role from machine operator to experiment architect.

\section*{Methods}

The Benchmarking Autonomy in Scientific Experiments (BASE) Scale was developed through a comparative analysis of existing autonomous standards in the automotive and microscopy sectors. We identified a functional misalignment between these owner-operator models, which assume extended instrument access for agent training, and the operational constraints of large-scale user facilities. Specifically, we isolated the transient user constraint, defined by limited beamtime allocations of typically less than 48-96 hours, which renders reinforcement learning approaches requiring on-instrument training impractical.

The taxonomy levels were derived by mapping the agency definitions of the SAE J3016 standard to scientific workflows~\cite{on2021taxonomy}. We defined the boundaries between levels by identifying two critical discontinuities in the experimental feedback loop. The first discontinuity is the Inference Barrier at Level 3, defined by the computational latency required to invert raw detector data into a semantic 3D space. The second is the Liability Threshold at Level 4, defined by the shift in risk ownership from the human operator to the validation framework when an agent is granted active hardware control. To quantify the agency of these levels, we integrated information theory principles and defined Entropy-Scaled Measurement Efficiency ($E_{SME}$) as the primary metric for differentiating semantic agents from passive automation. The mathematical derivation of this metric is detailed in our companion paper~\cite{lu2025autonomousbatteryresearchprinciples}.

To illustrate the practical utility of the taxonomy, we present the degradation of high-nickel battery cathodes as a demonstrator case study. We apply the BASE framework to estimate the theoretical efficiency limits of a scripted Level 1 approach versus a semantic Level 3 approach. By establishing that failure in this system is driven by rare geometric features, such as high kurtosis, we demonstrate that a semantic agent targeting information-rich voxels yields a theoretical order-of-magnitude improvement in efficiency compared to standard grid-based automation.

\section*{Acknowledgements}

JLH acknowledges funding from the Rutherford Appleton Laboratory and The Faraday Institution through the Emerging Leader Fellowship (FIELF001), and from Research England's `Expanding Excellence in England' (E3) fund via the ``Multi-scale Multi-disciplinary Modelling for Impact'' program (M$^3$4Impact).

\section*{Author contributions statement}
J.L.H. is the sole author. He conceived the study, developed the BASE taxonomy, and wrote the manuscript.

\section*{Competing interests}
The author declares no competing interests.

\bibliography{sample}

\end{document}